# Bidirectional Soliton Rain Dynamics Induced by Casimir-Like Interactions in a Graphene Mode-Locked Fiber Laser

Kfir Sulimany[1], Ohad Lib[1], Gilad Masri[3], Avi Klein[3], Moti Fridman[3], Philippe Grelu[2], Omri Gat[1], and Hadar Steinberg[1],

*(1) Racah Institute of Physics, The Hebrew University of Jerusalem, Jerusalem 91904, Israel*
*(2) Laboratoire ICB, UMR 6303 CNRS, Université Bourgogne-Franche-Comté, 9 avenue A. Savary, 21078 Dijon, France*
*(3) Faculty of Engineering and the Institute of Nanotechnology and Advanced Materials, Bar-Ilan University, Ramat Gan 5290002, Israel*

**Abstract**:
We study experimentally and theoretically the interactions among ultrashort optical pulses in the *soliton rain* multiple-pulse dynamics of a fiber laser. The laser is mode-locked by a graphene saturable absorber fabricated using the mechanical transfer technique. Dissipative optical solitons aggregate into pulse bunches that exhibit complex behavior, which includes acceleration and bi-directional motion in the moving reference frame. The drift speed and direction depend on the bunch size and relative location in the cavity, punctuated by abrupt changes under bunch collisions. We model the main effects using the recently proposed noise-mediated pulse interaction mechanism, and obtain a good agreement with experiments. This highlights the major role of long-range Casimir-like interactions over dynamical pattern formations within ultrafast lasers.

Dissipative solitons are localized waves in open systems far from equilibrium, whose existence results from a balance of dissipative and dispersive effects. They appear in numerous physical areas including reaction-diffusion systems, neurological and ecological sciences, fluid dynamics, and photonics [1,2]. In photonics, spatial dissipative solitons are stabilized by dynamical attractors and evolve like discrete particles in effective media [3]. A major challenge in their study is the distillation of an effective low-dimensional dynamical system governing pulse position and speed [4], which would determine the temporal evolution of multiple-pulse patterns. This implies an effective modeling of pulse interactions whose range and complexity are determined by the physical system. Mode-locked fiber lasers exhibit an extensive pallet of short and long-range pulse interactions. The former take place when pulse tails overlap [5-8], and the latter when pulses interact over separations orders of magnitude beyond their individual extension – and are hence mediated. In fiber lasers, where slow gain depletion and recovery dynamics create an effective long-range repulsive force [9] solitons pulses distribute equally along the cavity (harmonic mode-locking) [10]. Other interactions can be mediated by perturbations of an extended background field or propagation medium (e.g. electrostriction [11,12]).

Pulse interactions and soliton molecule formation in fiber lasers for long hinted at the existence of long-range attractive interactions, mediated by a background cavity field [8,10,13]. Recently, an interaction mechanism combining the effect of gain depletion in the presence of a noisy quasi-continuous wave permeating the laser cavity was discovered [14]. The interaction results from reduced field fluctuations consecutive to gain depletion, following the passage of a pulse. Fluctuations reduction decreases the pulse temporal jitter, thereby biasing it and setting a pulse motion. This noise-mediated interaction (NMI) mechanism shares some intriguing properties with the Casimir effect in quantum electrodynamics [15], where macroscopic objects experience an effective interaction as a consequence of the suppression electromagnetic field fluctuations. In NMI (although essentially a nonequilibrium effect), and the Casimir effect, distant objects inhibit microscopic fluctuations in extended electromagnetic modes. The consequent breaking of spatio-temporal homogeneity gives rise to weak interactions between these objects. Fortunately, in ultrafast laser systems the NMI mechanism is readily observable despite its relative weakness, which makes it work on the convenient second time scale. In the optical setting, NMI naturally combine with other interactions to create interesting stationary and variable pulse configurations, making it an excellent platform to study Casimir-like forces.

Finite gain relaxation time leads to a dynamical system with broken parity symmetry, resulting in Aristotelian forces where the action-reaction principle is violated. This was underlined in the different physical context of a cavity-extended semiconductor laser system [16], where the gain relaxation time is much shorter than the cavity roundtrip time, long-range interactions are absent, and dissipative solitons qualify as individually addressable localized structures [17]. In contrast, within the fiber laser platform, the long-range interactions among dissipative solitons lead to self-organized pattern formations, stationary as well as dynamic ones, which thrive over the entire laser cavity.

Soliton rain (SR) is dramatic multiple-pulse phenomenon, allowing processes of soliton pulse formation, drift, and collision to be followed in real time on an oscilloscope monitor. The SR dynamics takes place when multiple solitons interact in the presence of a noisy continuum field whose time-averaged power is non-uniform. In the strong continuum region, large fluctuations trigger the formation of new solitons that drift toward a compact soliton bunch – dubbed as the condensed soliton phase, which straddles a strong-weak continuum transition. The condensed soliton phase radiates energy, conserving an average number of constituents. SR displays a pattern of drift, collision and radiation, analogous to the cycle of water in the natural environment. It was discovered experimentally in a fiber laser employing nonlinear polarization evolution, a method producing an effective saturable absorber (SA) effect [18,19], and its universal dynamic features confirmed using other cavity architectures [20,21] and different SA types, including graphene [22]. Until now, however, there has not been any pulse interaction model explaining the dramatic soliton drift motions observed in the SR dynamics.

In this work we study the implications of the NMI theory in the SR dynamics using experimental, analytical and numerical approaches. We find the SR regime as being the most appropriate dynamical regime able to reveal, for the first time, such dramatic importance of Casimir-like interaction among a large variety of dissipative optical structures. We construct an all-fiber laser cavity, easily driven into the many-pulse regime at low pump intensity by employing a graphene SA requiring a low saturation power [23]. From the laser output, we record a wide range of experimental data, modelled using an analytic expression based on NMI theory. Numerical simulations, tracking soliton drift motions, exhibit good qualitative agreement with the experiment, and assert NMI as a dominant mechanism in long-range pulse interactions in fiber lasers.

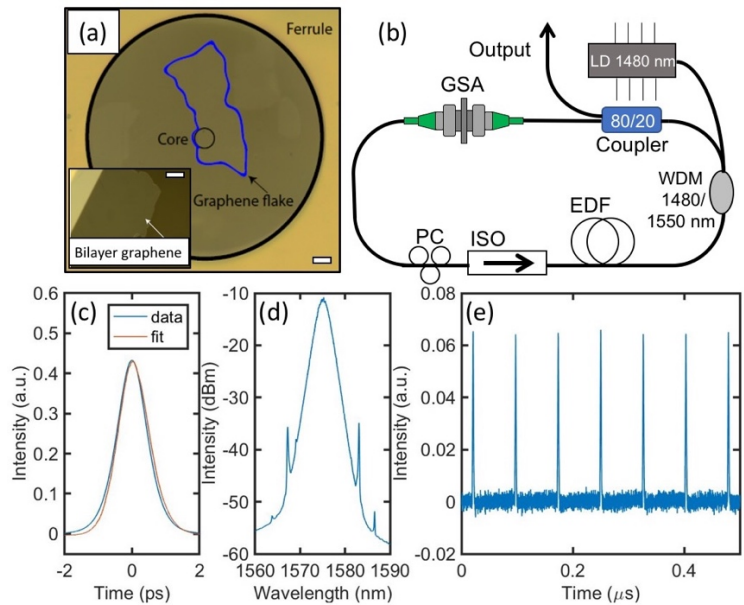

Figure 1: Experimental environment. (a) Bilayer graphene flake following transfer on the top of the fiber facet, covering the entire core, forming a graphene-based saturable absorber (scale bar 10 µm). Inset: The same graphene flake before transfer, on top of PDMS (scale bar 20 µm). (b) The ultrafast fiber laser setup, integrating the graphene saturable absorber (GSA). See text for abbreviations. (c) Autocorrelation trace, 1.1 ps wide and approximately hyperbolic-secant shaped (fit). (d) Optical spectrum, and (e) oscilloscope trace of the pulse train.

The realization of graphene SA for low-power fiber lasers involves the placement of graphene flakes on the fiber core, achievable using methods such as liquid phase exfoliation [24-26], chemical vapor deposition [27-29], carbon segregation [30], and micro-mechanical cleavage [31,32]. Here we developed a novel method to deposit clean graphene flakes on the core of an optical fiber, utilizing the mechanical transfer technique [33] which involves exfoliation of van der Waals materials on a sacrificial substrate, and their subsequent placement at desired locations. Graphene is first exfoliated on top of polydimethylsiloxane (PDMS), where it is identified and characterized. We then place the graphene precisely on top of the optical fiber core. This method allows on-demand positioning of high quality graphene flakes of desired shape and size, while retaining their crystallographic integrity. We expect it to find multiple applications in fiber-integrated 2D structures. This is shown in Fig. 1(a), where the inset depicts a microscope image of a bilayer graphene flake on PDMS. The flake is transferred onto an angled-physical-contact ferrule connector (FC-APC) or a bare fiber adapter connected with a fiber FC adapter. The main panel presents an image of the graphene flake covering the entire fiber core (blue outline).

The laser setup (Fig. 1(b)) is an all-fiber integrated ring cavity, 15.3-meter long, utilizing a 10-meter Erbium-doped fiber (EDF) gain medium. The EDF is core-pumped by a 1480-nm diode laser through a wavelength-division multiplexer (WDM). A polarization-independent optical isolator (ISO) ensures unidirectional lasing, and a fiber polarization controller (PC) tunes the overall low cavity birefringence. The GSA is sandwiched between the APC connectors of single-mode fiber jumpers, and a 20/80 coupler provides the laser output. The cavity dispersion is anomalous, at -28 ps$^2$. The mode-locking operation self-starts at a threshold pump power $P = 43\ mW$, and is maintained up to $P = 300\ mW$. The output power scales almost linearly with the pump power above threshold, with a maximum of 32 mW at $P = 300\ mW$. The ultrafast laser output is analyzed with an optical spectrum analyzer, a multi-shot optical autocorrelator based on second-harmonic generation, and a fast photodiode connected to a real-time oscilloscope. The autocorrelation trace of the mode-locked pulses after a ~50 cm single mode fiber lead is shown in Fig. 1(c). Assuming a squared hyperbolic-secant (*sech*) temporal power profile, the deconvolution indicates a pulse duration of ~1.1 ps. The time-bandwidth product is 0.33, very close to the Fourier-transform limit of 0.315 for *sech* pulses, confirming the major role of the soliton pulse shaping in this regime. A typical mode-locked output spectrum is displayed in Fig. 1(d). The central wavelength is 1575 nm, and the full width at half maximum (FWHM) bandwidth is ~2.5 nm. In the time domain trace, Fig. 1(e), the pulse train period is 76.5 ns, as expected from the fundamental repetition frequency of a 15.3 m-long fiber cavity.

For laser cavities endowed with an anomalous dispersion, soliton pulse shaping leads to a pulse energy limitation that mainly scales with the ratio between the cavity dispersion and the fiber effective nonlinearity. We estimate the intracavity single-pulse energy and peak power to be ~50 pJ and ~40 W, respectively. Consequently, beyond a minimal pumping level, single-pulse operation becomes unstable, leading to nucleation of additional pulses [10,34-36]. Relatively low pulse energy favors multiple pulsing with large pulse numbers, a prerequisite to SR dynamics. Whereas stationary multi-pulse patterns are commonly produced, ranging from loose pulse bunches to compact soliton molecules [10,36], higher pumping is liable to trigger dynamical instabilities leading to non-stationary states, from pulsating to chaotic ones [18,19,37-41]. Nevertheless, within a large range of parameters, solitonic pulse shaping dominates, allowing the complex waveforms to be well described as superpositions of identical moving pulses on a weak inhomogeneous and noisy background, as is the case within SR dynamics.

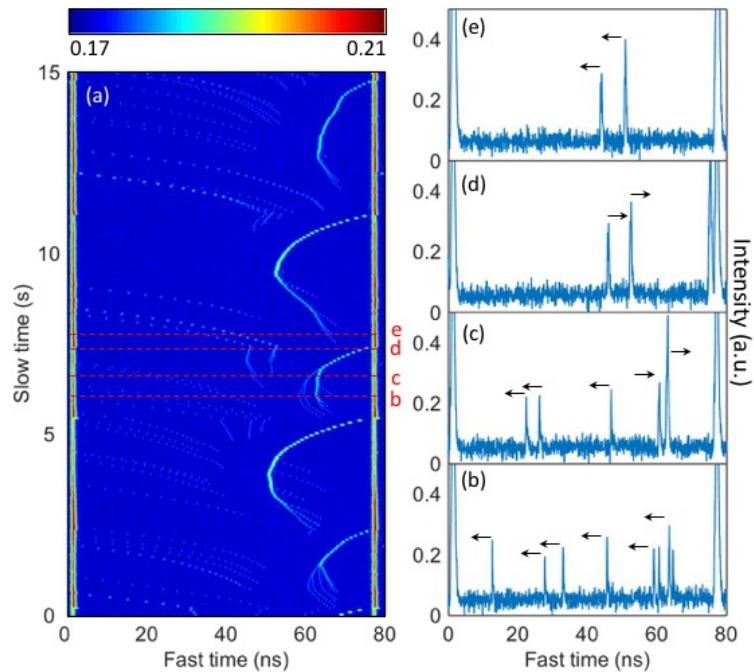

Figure 2: Experimental recording of accelerated soliton dynamics. (a) Slow time vs. fast time plot, optical power color-coded. Color code chosen to create contrast among smaller bunches. The two strong peaks near the edges are the condensed soliton

phase anchor repeated after one cavity roundtrip (76.5 ns). Trajectories exhibit bi-directional accelerated motion relative to the anchor. The red dashed lines mark the times of individual oscilloscope traces displayed in panels (b-e) from bottom to top, with an arrow on top of each pulse showing its relative direction of motion.

The SR dynamics is presented in Fig. 2. Panel (a) is a spatiotemporal representation combining a succession of numerous oscilloscope traces recorded over 15 seconds. The intense line on the left represents the condensed soliton phase, which repeats after a single cavity roundtrip of 76.5 ns. It serves as the oscilloscope fast time trigger reference, and an anchor for the soliton trajectories. The slow time is the actual time passing over successive roundtrips. The rich soliton dynamics reveals itself in the form of complex pulse trajectories relative to the anchor. Soliton pulse bunches form in regions of strong continuum, and subsequently drift and collide with other pulses until they reach the condensed soliton phase. Between such merge effects, the anchor energy appears to reduce, in analogy to the evaporation of a condensed phase [21]. Note that the structure of soliton bunches is not resolved in the time domain, owing to the limited oscilloscope bandwidth. Consequently, pulse bunches appear as large pulses on the oscilloscope trace, whose amplitude is proportional to the number of pulse constituents, all pulses having the same energy [21]. Particular traces corresponding to the red dashed horizontal lines are shown in Panels (b-e), an arrow attached to each pulse indicating its relative direction in the moving reference frame of the anchor.

The pulse trajectories depicted in Fig. 2(a) have the following characteristics: (i) Soliton bunches propagate either faster (left) or slower (right) than the anchor. (ii) Soliton bunches accelerate. (iii) The drift velocity depends on the soliton-bunch power. (iv) When soliton bunches merge, the relative velocity of distant pulses is affected. For example, on the right side of Fig. 2(a), at the time interval between the snapshots (b) and (c), we observe the merging of soliton bunches. The resulting stronger bunches change direction, merge with each other, and finally merge back into the anchor. This cavity hence exhibits SR dynamics that differ from those previously reported [18-22], and we refer to it as a *bi-directional* SR. The most likely enabling factors for observing this regime experimentally are the low saturation power of the few-atom thick graphene layer [42], compatible with a low dissipative soliton energy scale, and the stronger effects of noise and continuum that underlie the SR state, owing to the few-percent contrast of the nonlinear transfer function of the SA.

Our theory of pulse motion is based on the NMI mechanism developed in [14], reminiscent of the Casimir effect in quantum electrodynamics [15]. The pulse interaction arises as a by-product of the pulse timing jitter caused by the nonlinear overlap interaction between the pulses and the random quasi-CW floor [11,43]. As shown in Fig. 3(a), gain is depleted by the pulses. The quasi-CW floor, proportional to the gain, is thus inhomogeneous, its intensity is sharply reduced after a pulse passage, recovering gradually afterwards. Since the interaction with the quasi-CW floor makes the pulses diffuse [44], the diffusion strength proportional to its intensity, the pulse diffusion will depend on the pulse position within the waveform, mathematically similar to a Brownian motion in an inhomogeneous environment, the bias translating into the drift velocity of the pulses. Specifically, denoting the timing of the *n*-th pulse by $t_n$, its diffusion constant by $D_n$, and the slow time coordinate or propagation distance by $z$, then

$$\frac{\partial \langle t_n \rangle}{\partial z} = \frac{1}{2} \frac{\partial D_n}{\partial t_n} \qquad (1)$$

Where the brackets stand for noise averaging. The diffusion constant $D_n$ is proportional to the local intensity of the quasi-CW, in turn determined by the local net gain as [14]:

$$D_n(t_n) \propto \left( \frac{1}{\sqrt{l-g(t_n^-)}} + \frac{1}{\sqrt{l-g(t_n^+)}} \right), \qquad (2)$$

where $l$ is the total small signal loss. Eqs. 1 and 2 dictate the relative motion of soliton bunches in relation to the noise background expressed by the local gain. As shown below, bunches drift in the direction of reduced noise power. This is analogous to the Casimir effect, where macroscopic object experience a pull towards a region in space where vacuum fluctuations are suppressed (noting that here the noise does not consist of zero point fluctuations.)

The variations in $g$ are small compared with the mean gain $\bar{g}$, but may induce large changes in the net loss. Since the gain is depleted by the pulses, whose temporal width is much smaller than any other relevant time-scale, the depletion is viewed as instantaneous - $g(t_n^-)$ and $g(t_n^+)$ denoting the gain coefficients before and after the *n*-th pulse, respectively. Between pulses, the gain recovers according to:

$$\frac{dg}{dt} = \frac{g_u - g}{t_{rec}} \quad (3)$$

where $g_u$ is the unsaturated gain and $t_{rec}$ is the recovery time; these parameters are determined by the pumping regime. Since gain variations are much smaller than $g_u - \bar{g}$, the recovery rate is very close to the constant $(g_u - \bar{g})/t_{rec}$, and the gain profile is a sawtooth-shaped function with downward jumps at each pulse, and a constant-slope linear rise between the pulses.

Fig. 3 demonstrates the predictive power of the dynamical system described by Eqs. (1–3), solved numerically. Here, insight can be gained by following the temporal evolution of the noise floor in presence of pulse bunches of different energy. The relative pulse velocity, determined by Eq. (1), becomes highly sensitive to gain variation where the gain *g* comes close to the loss *l*, which is where both $D_n$ and its time-derivative are greatest. As a result, the energetic bunches propagate at a different relative velocity with respect to smaller pulses in their wake, where the noise-background is strongly suppressed. In the anchor reference frame, a small bunch following a large one will appear to fall towards it. As the small pulse comes closer to the anchor, it experiences even lower gain levels, causing it to further accelerate. This acceleration is clearly seen in the simulated pulse trajectory (red) in Fig. 3(a).

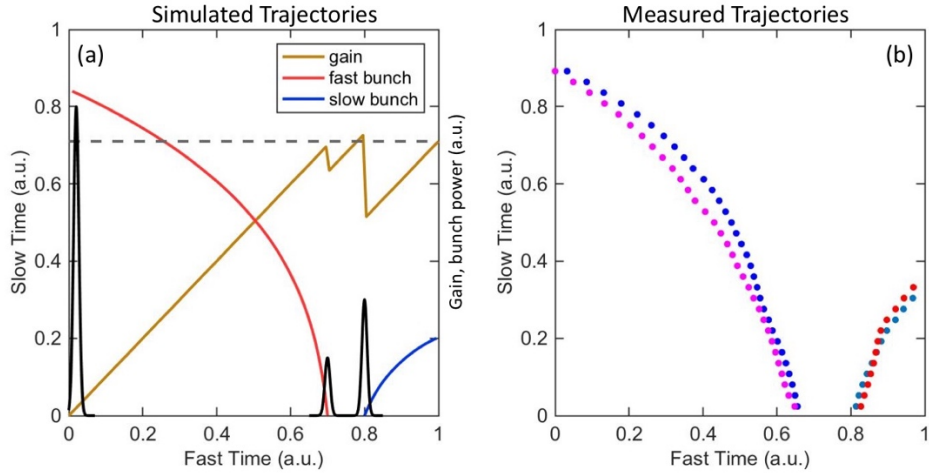

Figure 3: Analysis of accelerated pulse dynamics. (a) Simulation of pulse-bunch trajectories located to the right (blue curve) and left (red curve) of the stagnation point. A schematic illustration of NMI mechanism is overlaid, showing the pulse bunches in black and the gain level in orange (not to scale). The stagnation point is found where gain levels reach the gain value before the anchor (horizontal dashed line). (b) Typical experimental traces of pulse bunch trajectories, measured in the fast time scale, as a function of slow time evolution (extracted at different initial times from Fig. 2(a)). The slow time is relative to each initial motion start time (fluctuations in the trajectories at early slow times caused by pulse collisions).

The temporal evolution of the gain reveals another important feature of the pulse dynamics: At the point where the gain recovers to its value at the anchor, shown by a dotted line in Fig. 3(a), Eq. 2 dictates that the bunch drift velocity is approximately equal to the anchor velocity. Since beyond this point the bunch is faster than the anchor, and before it the bunch is slower, the gain parity point is an unstable stagnation point of the pulse flow, in agreement with simulation results (blue trajectory in Fig. 3(a)). Thus, simulated trajectories agree well with the experimental ones, presented in Fig. 3(b), depicting superimposed four experimental single-pulse trajectories in the anchor reference frame. In agreement with the theory, each small bunch approaches the anchor by either moving forward or backward, depending on its initial position inside the cavity, the domains of forward- and backward-motion separated by the unstable stagnation point.

Fig. 4 compares experiments with NMI simulations in the presence of numerous pulse bunches. In the simulation, the initial condition consists of an arbitrary number of pulse bunches, each composed of an arbitrary number of solitons. The experimental plots (a,c) are characterized by abrupt changes in the pulse direction as a response to pulse merging events. For example, the pulses marked by red dots in Panel (a) merge and then drift rightwards until reaching the anchor at the slow-time marked as "$z_1$". At the same instant, a second set of pulses, marked in yellow, change their direction. This effect, too, is understood when considering the noise profile, which changes as the energy of the "red" pulse is added to the anchor: a more energetic anchor shifts the position of the stagnation point to the right. A pulse which, at $z < z_1$, was at the right of the stagnation point and hence a right-mover, abruptly becomes placed left of the stagnation point, and hence a left mover. This process is reversed as the anchor loses energy over time. All these major observed effects – pulse acceleration, bi-directionality, and changes of direction - are explained by the NMI model, and qualitatively well reproduced by the numerical simulations.

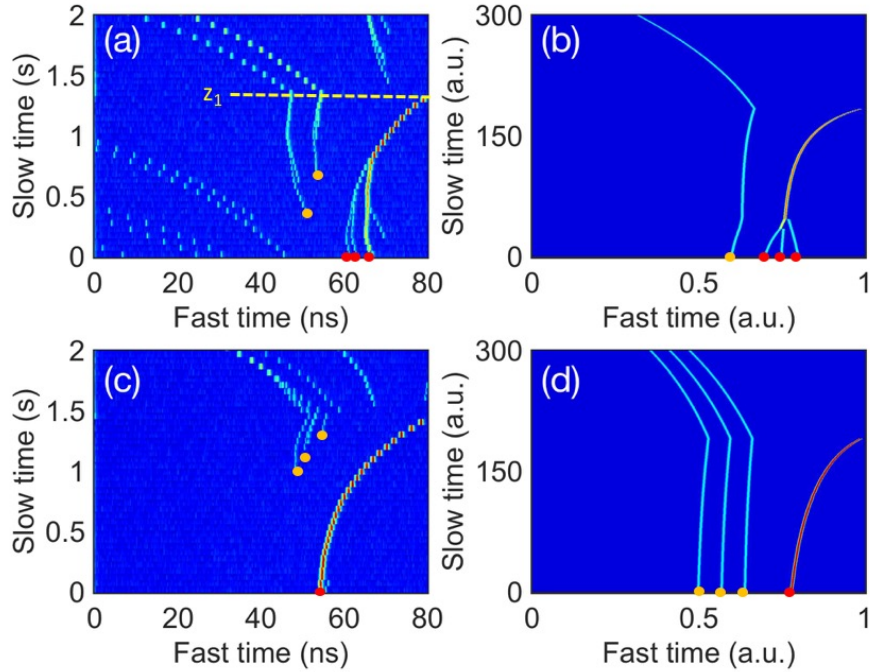

Figure 4: Complex many-pulse trajectories, displayed as color-coded oscilloscope readout vs. fast and slow times, comparing the experiments (left panels a,c) with simulations (right panels b,d). Simulations are based on the NMI theory, assuming the annihilation of a fixed fraction of bunch energy during collisions. The anchor is positioned at the edges of the waveform. The pulses are marked either at zero slow time or at the time of their formation, with red or yellow points, respectively.

In summary, we model dissipative soliton trajectories under long-range many-body interactions. Such theory is long-sought, as pulses in the SR dynamical regime sequentially interact through different and competing forces, involving times scales spanning 12 orders of magnitude. This theory represents, to our knowledge, the only example where solitons interact by suppression of fluctuations, akin to the Casimir interaction. While a comprehensive SR theory accounting for stochastic soliton formation and annihilation is beyond the current scope, the quantitative success of the NMI theory should impact future research of soliton pattern formations in ultrafast lasers. Ultimately, if such dynamics could be scaled down to ultralow powers in miniature high-Q laser cavities, so as to produce pulses weak enough to be sensitive to single photon fluctuations, then the NMI could be used to sense the quantum fluctuations of the cavity electromagnetic field.

## *ACKNOWLEDGEMENTS*

H.S. acknowledges support from Israeli Science Foundation Grant No. 1363/15. Ph.G. is supported from ANR LABEX Action and FEDER funds.